\documentclass[runningheads]{llncs}
\usepackage{graphicx}
\usepackage{hyperref}
\hypersetup{
    colorlinks=true,
    linkcolor=blue,
    citecolor=blue,
    urlcolor=blue
    }

\usepackage[utf8]{inputenc}
\usepackage{caption}
\usepackage{subcaption}
\usepackage{booktabs}
\usepackage{multirow}
\usepackage{color}

\usepackage{tikz}
\usetikzlibrary{arrows,backgrounds}

\usepackage{mwe}
\usepackage{lipsum}

\usepackage{amsmath}
\usepackage{amssymb}
\usepackage{mathtools}

\DeclareMathOperator*{\argmin}{arg\,min}

\newcommand{\comment}[1]{}

\begin{document}
\title{Post-Error Correction for Quantum Annealing Processor using Reinforcement Learning}
\titlerunning{Post-Error Correction for  Quantum Annealing Processor using RL}
%
\author{Tomasz Śmierzchalski\inst{1,2} \and
Łukasz Pawela\inst{2}\and Zbigniew Puchała\inst{2,3} \and 
Tomasz Trzcinski\inst{1} \and Bartłomiej Gardas\inst{2}}
\authorrunning{T. Śmierzchalski et al.}
%
\institute{
Warsaw University of Technology,
Pl. Politechniki 1,
00-661 Warsaw, Poland \and
Institute of Theoretical and Applied Informatics, Polish Academy of Sciences, Bałtycka 5, 44-100 Gliwice, Poland
\and
Faculty of Physics, Astronomy and Applied Computer Science, Jagiellonian~University, 30-348 Krak\'{o}w, Poland
}
\maketitle              

\begin{abstract}
Finding the~ground state of the~Ising spin-glass is an important and challenging problem (NP-hard, in fact) in condensed matter physics. However, its applications spread far beyond physic due to its deep relation to various combinatorial optimization problems, such as travelling salesman or protein folding. Sophisticated and promising new methods for solving Ising instances rely on quantum resources. In particular, quantum annealing is a quantum computation paradigm, that is especially well suited for Quadratic Unconstrained Binary Optimization (QUBO). Nevertheless, commercially available quantum annealers (i.e., D-Wave) are prone to various errors, and their ability to find low energetic states (corresponding to solutions of superior quality) is limited. This naturally calls for a post-processing procedure to correct errors (capable of lowering the~energy found by the~annealer). As a proof-of-concept, this work combines the~recent ideas revolving around the~DIRAC architecture with the~Chimera topology and applies them in a real-world setting as an error-correcting scheme for quantum annealers. Our preliminary results show how to correct states output by quantum annealers using reinforcement learning. Such an approach exhibits excellent scalability, as it can be trained on small instances and deployed for large ones. However, its performance on the chimera graph is still inferior to a typical algorithm one could incorporate in this context, e.g., simulated annealing.

\end{abstract}

\section{Introduction}

Many complex and significant optimization problems (such as all of Karp's 21 NP-complete problems~\cite{L2014}, the~travelling salesman problem~\cite{KT1985}, the~protein folding problem~\cite{BW1987}, financial portfolio management~\cite{quantumfinance}) can be mapped into the problem of finding the~ground state of the~Ising spin-glass. Sophisticated and promising new methods for solving Ising instances rely on quantum computation, particularly quantum annealing.

Quantum annealing is a form of quantum computing particularly well-tailored for optimization~\cite{KN1998,SMTC2002}. It is closely related to adiabatic quantum computation~\cite{M2014}, a paradigm of universal quantum computation which relies on the~adiabatic theorem~\cite{K1950} to perform calculations. It is equivalent (up to polynomial overhead) to the~better-known gate model of quantum computation~\cite{M2014}. Nevertheless, commercially available quantum annealers (i.e., D-Wave) are prone to various errors, and their ability to find low energetic states is limited.

Inspired by the~recently proposed deep reinforcement learning method for finding spin glass ground states~\cite{FSNLSL2021}, here, we propose a new post-processing error correction schema for quantum annealers called Simulated Annealing with Reinforcement (SAwR). In this procedure, we combine deep reinforcement learning with simulated annealing. We employ a graph neural network to encode the~Ising instance into an ensemble of low-dimensional vectors used for reinforcement learning. The~agent learns a strategy for improving (finding a lower energy state) solutions given by the~physical quantum annealer. The process of finding the lower energy state involves "flipping" spins one by one according to the learned strategy and recording the energy state after each step. The solution is defined as the lowest energy state found during this procedure. In Simulated Annealing with Reinforcement, we start with simulated annealing and, at low temperature, we replace the~Metropolis-Hasting criterion with a single pass of spin flipping procedure. 

Unlike recent error-correcting schema~\cite{PAL2014,PAL2015,V2015} we do not utilize multiple physical qubits for representing single logical qubits. This approach allows for a far greater size of problems to which our method applies. However, the performance of SAwR is still inferior to a typical algorithm one could incorporate in this context, such as simulated annealing. Nevertheless, using reinforcement learning for post-error correction is still an open and promising avenue of research.

\begin{figure}[ht!]
    \centering
    \includegraphics[width = \textwidth]{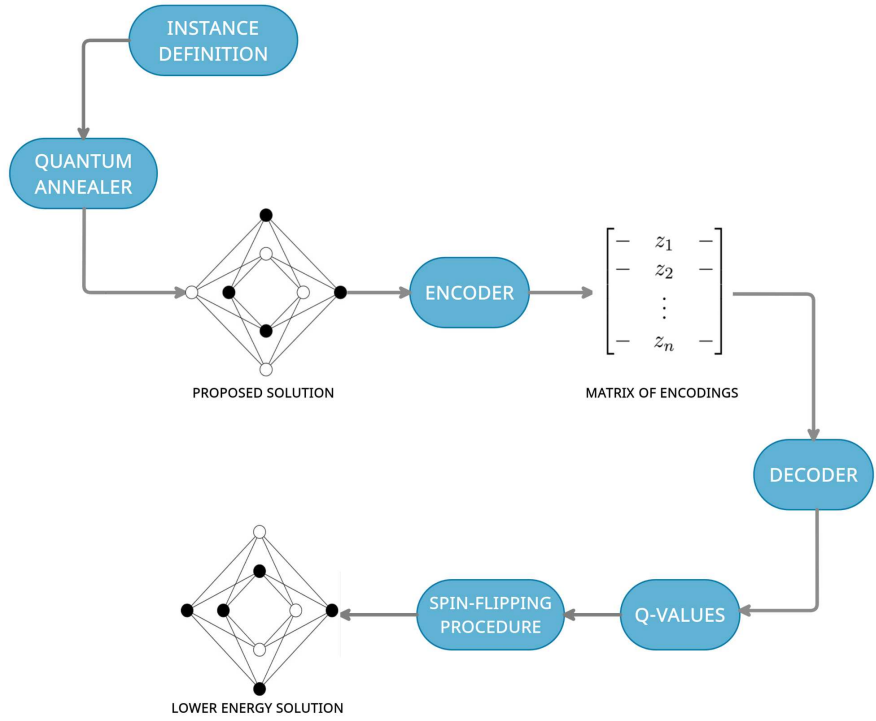}
    \caption{\small Overview of our method, arrows represent subsequent steps. First, we define the Ising instance by providing couplings strength and biases (external magnetic field strength). Then we obtain the proposed solution from a quantum annealer. Here, it is represented by a single Chimera unit cell rendered as a graph. Black nodes represent spin value $\sigma_i = -1$ and white nodes represent spin value $\sigma_i = 1$. Edges represent couplings between spins. In the next step, we encode such an instance using a graph neural network into the matrix of encodings, where each row corresponds to the embedding of a vertex. Then this matrix is passed through a decoder to obtain Q-values of actions associated with each vertex. The spin flipping procedure involves "flipping" spins one by one according to Q-Values, starting from the highest and recording the energy state after each step. The solution is defined as the lowest energy state found during this procedure.}
    \label{schema}
\end{figure}

\section{Ising Spin Glass and Quantum Annealing}

\subsection{Quantum Annealing in D-Wave }

The Ising problem is defined on some arbitrary simple graph $G$. An Ising Hamiltonian is given by:

\begin{equation}
    \label{ising}
    H_{\text{Ising}}(\sigma) = \sum_{\langle i,j \rangle \in G} J_{i,j} \sigma_i \sigma_j + \sum_{i} h_i \sigma_i,
\end{equation}

\noindent where $\sigma_i \in \{+1,-1\}$ denotes the~$i$-th Ising spin, $\sigma = \{ \sigma_1, \ldots, \sigma_n \}$ is the~vector representation of a spin configuration. $\langle i,j \rangle$ denotes neighbours in graph $G$, $J_{i,j}$ strength of interaction (coupling coefficient) between $i$-th and $j$-th spin. $h_i$ is external magnetic field (bias) affecting $i$-th spin. The~goal is to find spin configuration, called ground state,  

\begin{equation}
\label{goal}
    \sigma^{*} = \argmin_{\sigma} \, H_{\text{Ising}}(\sigma).
\end{equation}

\noindent such that energy of $H_{\text{Ising}}$ is minimal. This is an NP-hard problem. Quantum annealing is a method for finding the~ground state of (\ref{ising}). This is done by adiabatic evolution from the initial Hamiltonian $H_X$ of the~Transverse-field Ising model to the final Hamiltonian $H_{\text{Ising}}$. The~Hamiltonian of this process is described by

\begin{equation}
    \mathcal{H}(t) = A(t)H_{X} + B(t)H_{\text{Ising}},
\end{equation}

\noindent where $H_X = \sum_i \hat{\sigma_i}^x$ and $\hat{\sigma_i}^x$ is the~standard Pauli $X$ matrix acting on the~$i$-th qubit. The~function $A(t)$ decreases monotonically to zero, while $B(t)$ increases monotonically from zero, with $t \in [0, t_f]$, where $t_f$ denotes total time of anneal~\cite{JRS2007,SMTC2002}. For a closed system, the~adiabatic theorem guarantees that if the~initial state is the~ground state, then the~final state will be arbitrarily close to the~ground state, provided all technical requirements are met~\cite{K1950}. In summary, the~systems start with a set of qubits, each in a superposition state of $-1$ and $1$. By annealing, the~system collapses into the~classical state that represents the~minimum energy state of the~problem, or one very close to it. 

The D-Wave 2000Q is a psychical realization of the~quantum annealing algorithm. Sadly, the~idealized conditions of the~adiabatic theorem are nearly impossible to be realized in a physical device. In such an open system, there is inevitable thermal noise which may cause decoherence~\cite{decoherence}. Furthermore, due to technical limitations, those device suffers from programming control errors on the~$h_i$ and $J_{i,j}$ terms, which can unintentionally cause the~annealer to evolve according to the~wrong Hamiltonian~\cite{control}.

\subsection{D-Wave 2000Q}

\begin{figure}[t]
    \centering
    \includegraphics[scale = 0.34]{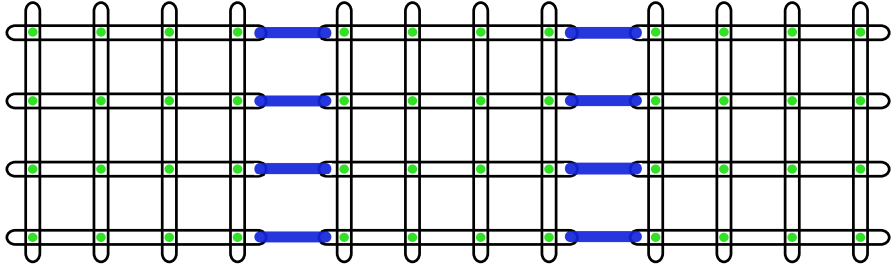}
    \caption{\small Qubits are represented as horizontal and vertical loops. This graphic shows three coupled unit cells with eight qubits each. Green dots represent internal couplers connecting qubits inside the~unit cell, while blue bars represent external couplers connecting different unit cells.}
    \label{qubits}
\end{figure}

At the~heart of every D-wave quantum annealer lies the~Quantum Processing Unit (QPU), which is a lattice of interconnected qubits. While its physical details are beyond the scope of this paper\footnote{Interested reader may find details in~\cite{dwave_IEEE} and~\cite{dwave_nature1} }, it is necessary to mention some technical details. In QPU, qubits can be thought of as loops being "oriented" vertically or horizontally (see figure \ref{qubits}) and connected to each other via devices called couplers. How qubits and couplers are interconnected is described by QPU topology. 

As of the~time of writing, there are two available architectures of D-Wave quantum annealers, namely 2000Q with Chimera topology deployed in 2017 and Advantage with Pegasus topology deployed in 2020. Third, called Advantage~2 with Zephyr topology~\cite{zephyr} is stated to release in 2023-2024~\cite{roadmap}. In this work, we will focus on the~2000Q device and Chimera topology. 

\subsubsection{Chimera Topology}

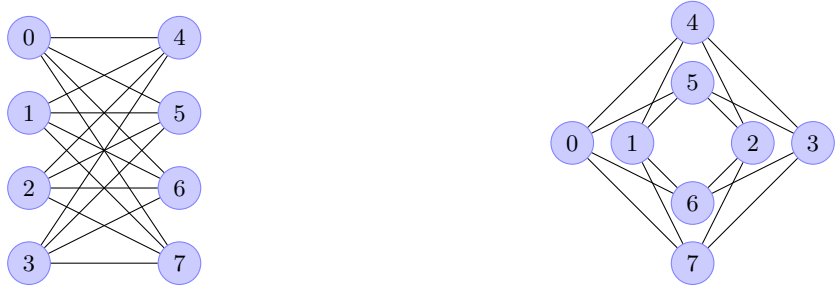
\begin{figure}
    \centering
    \begin{subfigure}[t]{0.3\textwidth}
        \centering
        \tikzstyle{node}=[circle,draw=blue!50,fill=blue!20]
        \begin{tikzpicture}
        
            \node at (0,0) [node] (1) {3};
            \node at (0,1) [node] (2) {2};
            \node at (0,2) [node] (3) {1};
            \node at (0,3) [node] (4) {0};

            \node at (2,0) [node] (5) {7}
            	edge (1)
                edge (2)
                edge (3)
                edge (4);
            \node at (2,1) [node] (6) {6}
            	edge (1)
                edge (2)
                edge (3)
                edge (4);
            \node at (2,2) [node] (7) {5}
            	edge (1)
                edge (2)
                edge (3)
                edge (4);
            \node at (2,3) [node] (9) {4}
            	edge (1)
                edge (2)
                edge (3)
                edge (4);
        \end{tikzpicture}
         \caption{\small Unit cell rendered as column}
         \label{column}
     \end{subfigure}
     \hfill
     \begin{subfigure}[t]{0.4\textwidth}
        \centering
        \tikzstyle{node}=[circle,draw=blue!50,fill=blue!20]
        \begin{tikzpicture}[scale=0.8]
        
        	\node at (-2,0) [node] (1) {0};
        	\node at (-1,0) [node] (2) {1};
            \node at (1,0) [node] (3) {2};
            \node at (2,0) [node] (4) {3};
            
            \node at (0,2) [node] (5) {4}
            	edge (1)
                edge (2)
                edge (3)
                edge (4);
            \node at (0,1) [node] (6) {5}
                edge (1)
                edge (2)
                edge (3)
                edge (4);
            \node at (0,-1) [node] (7) {6}
                edge (1)
                edge (2)
                edge (3)
                edge (4);
            \node at (0,-2) [node] (8) {7}
                edge (1)
                edge (2)
                edge (3)
                edge (4);

        \end{tikzpicture}    
        \caption{\small Unit cell rendered as cross}
        \label{cross}
     \end{subfigure}
     \hfill
    \caption{\small Different renderings of Chimera unit cell as graph. Nodes represent qubits and edges represent internal couplers.}
    \label{unit_cell}
\end{figure}

The basic building block of Chimera topology is a set of connected qubits called a unit cell. Each unit cell consists of four horizontal qubits connected to four vertical qubits via couplers which form bipartite connectivity as seen in figure \ref{unit_cell}. Unit cells are tiled vertically and horizontally with adjacent qubits connected, creating a lattice of sparsely connected qubits.

It is conceptually valuable to categorize couplers into \emph{internal couplers} which connect intersecting (orthogonal) qubits and \emph{external couplers} which connect colinear pairs of qubits (that is, pairs of qubits that lie in the~same row or column). The~notation $C_n$ describes the~Chimera grid composed of $n \times n$ coupled unit cells, consisting of $8n^2$ qubits. D-Wave 2000Q device is equipped with $C_{16}$ QPU, with more than 2000 qubits~\cite{manual}.

\section{Results}

\subsection{Reinforcement Learning Formulation}

We will consider standard reinforcement learning setting defined as Markov Decision Process~\cite{SB2018} where an agent interacts with an environment over a number of discrete time steps $t = 0,1, \ldots, T$. At each time step $t$, the~agent receives a state $s_t$ and selects an action $a_t$ from some set of possible actions $\mathcal{A}$ according to its policy $\pi$, where $\pi$ is a mapping from set of states $\mathcal{S}$ to set of actions $\mathcal{A}$. In return, the~agent receives a scalar reward $r_t$ and moves to next state $s_{t+1}$. The~process continues until the~agent reaches a terminal state $s_T$ after which the~process restarts. We call one pass of such process an episode. The~return  at time step $t$, denoted $R_t = \sum_{k=0}^{T-t} \gamma^k r_{t+k}$ is defined as sum of rewards that agent will receive for rest of the~episode discounted by discount factor $\gamma \in (0, 1]$. The~goal of the~agent is to maximize the~expected return from each state $s_t$.

We will start by defining state, action, and reward in the~context of the~Ising spin-glass model.

\begin{itemize}
\item \textbf{State}: a state $s$ represents the~observed spin glass instance, including both the~spin configuration $\sigma$, the~coupling strengths $\{ J_{ij} \}$ and values of external magnetic field $\{ h_i \}$. 

\item \textbf{Action}: an action $a^{(i)}$ means to flip spin $i$. By flipping spin we mean changing is value to opposite. For example, after agent performs action $a^{(i)}$, spin $\sigma_i = 1$ becomes $\sigma_i = -1$. Agent can flip each spin once.

\item \textbf{Reward}: the~reward $r(s_t; a^{(i)}_t; s_{t+1})$ is defined as the~energy change after flipping spin $i$ from state $s_t$ to a new state $s_{t+1}$.

\end{itemize}

Starting at $t = 0$, an agent flips one spin during each time step, which moves him to the~next state (different spin configuration). The~terminal state $s_T$ is met when the~agent has flipped each spin. The~solution is defined as spin configuration $\sigma$ corresponding lowest energy state found during this procedure.   

An action-value function $Q^{\pi}(s,a) = \mathbb{E}(R_t ~\vert~ s_t = s,~ a_t = a)$ is the~expected return for selecting action $a$ in state $s$ and following policy $\pi$. The~value $Q^{\pi}(s,a)$ is often called $Q$-value of action $a$ in state $s$. The~optimal action-value function $Q^{*}(s,a) = \max_{\pi} Q^{\pi}(s,a)$ which gives the~maximum action value for state $s$ and action $a$ achievable by any policy. As learning optimal action-value function is in practice infeasible, We seek to learn function approximator $Q(s,a;\Theta) \approx Q^{*}(s,a)$ where $\Theta$ is set of learnable model parameters. We denote policy used in such aproximation as $\pi_{\Theta}$.

\begin{figure}[t]
    \centering
    \includegraphics[scale = 0.6]{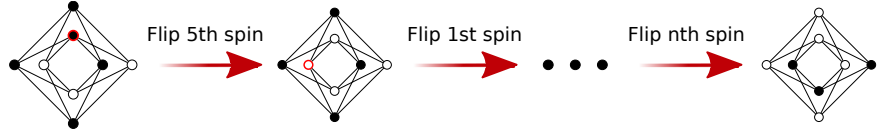}
    \caption{\small Overview of single episode. To simplify, we show it on a single Chimera unit cell. White nodes represent $\sigma_i = 1$ and black nodes represent $\sigma_i = -1$ We start in some state $s_0$ and one by one we flip spins until all spins are flipped. The~solution is defined as spin configuration $\sigma^*$ corresponding lowest energy state found during this procedure.}
    \label{fig:my_label}
\end{figure}


\subsection{Model Architecture}

Our model architecture is inspired by DIRAC (\textbf{D}eep reinforcement learning for sp\textbf{I}n-glass g\textbf{R}ound-st\textbf{A}te \textbf{C}alculation), an Encoder-Decoder architecture introduced in~\cite{FSNLSL2021}. It exploits the~fact that the~Ising spin-glass instance is wholly described by the~underlying graph. In this view, couplings $J_{i,j}$ become edge weights, external magnetic field $h_i$ and spin $\sigma_i$ become node weights. Employing DIRAC is a two-step process. At first, it encodes the~whole spin-glass instance such that every node is embedded into a low-dimensional vector, and then the~decoder leverages those embeddings to calculate the~$Q$-value of every possible action. Then, the agent chooses the~action with the~highest $Q$-value. In the~next sections, We will describe those steps in detail.

\subsubsection{Encoding}

 As described above, the~Ising spin-glass instance can be described in the~language of graph theory. It allows Us to employ graph neural networks~\cite{GRL,GSRVD2017}, which are neural networks designed to take graphs as inputs. We use modified SGNN (\textbf{S}pin \textbf{G}lass \textbf{N}eural \textbf{N}etwork)~\cite{FSNLSL2021} to obtain node embedding. To capture the~coupling strengths and external field strengths (i.e., edge weights $J_{i,j}$ and node weights $h_i$), which are crucial to determining the~spin glass ground states, SGNN performs two updates at each layer, specifically, the~edge-centric update and the~node-centric update, respectively.

Lets $z_{(i,j)}$ denote embedding of edge $(i,j)$ and $z_{(i)}$ embedding of node $i$. The~edge-centric update aggregates embedding vectors from from its adjacent nodes (i.e. for edge $(i,j)$ this update aggregate emmbedings $z_{(i)}$ and $z_{(j)}$), and then concatenates it with self-embedding $z_{(i,j)}$. Vector obtained in this way is then subject to non-linear transformation (ex. ReLU$(x)= \max(0,x)$). Mathematically it can be described by following equation

\begin{equation}
\label{edge}
    z_{(i,j)}^{k+1} = \text{ReLU}(\gamma_{\theta} (z_{(i,j)}^{k}) \oplus \phi_{\theta}(z_{(i)}^{k} + z_{(j)}^{k})),
\end{equation}

 \noindent where $z_{(i,j)}^{k}$ denotes encoding of edge $(i,j)$ obtained after $k$ layers. Similary $z_{(i)}^{k}$ denotes encoding of node $i$ obtained after $k$ layers, $\gamma_{\theta}$ and $\phi_{\theta}$ are some differentiable functions (ex. feed-foward neural networks) whih demends on set of paramethers $\theta$. Symbol $\oplus$ is used to denote concatenation operation. 
 
 The~node-centric update is defined in similar fashion. It aggregates embedding of adjacent edges, and then concatenates it with self-embedding $z_{(i)}$. Later we transform this concatenated vector to obtain final embedding. Using notation from equation \ref{edge}, final result is following:

\begin{equation}
    z_{(i)}^{k+1} = \text{ReLU}(\phi_\theta(z_{(i)}^k) \oplus \gamma_\theta(\text{E}_i^k)),
\end{equation}
\begin{equation}
    \text{E}_i^k = \sum_{j} z_{(i, j)}^k .
\end{equation}

\noindent Edge features are initialized as edge weights $\{J_{i,j}\}$. It is not trivial to find adequate node features, as node weights $\{h_i \}$ and spins ${\sigma_i}$ are not enough. 

It is worth noting that both those operations are \textit{message passing} schema~\cite{GRL}. Edge-centric update aggregate information about adjacent nodes of edge and edge itself and sends it as a "message" to this edge. Similarly, node-centric update aggregate information about edges adjacent to the~node and the~node itself. In those edges are also encoded information about neighbouring nodes.   

We also included pooling layers not presented in the~original design. We reasoned that after concatenation, vectors start becoming quite big, so we employ pooling layers to not only reduce the~model size but also preserve the~most essential parts of every vector.

As every node is a potential candidate for action, we call the~final encoding of node $i$ its \textit{action embedding} and denote it as $Z_i$. To represent the~whole Chimera (state of our environment), we use \textit{state embedding}, denoted as $Z_s$, which is the~sum over all node embedding vectors, which is a straightforward but empirically effective way for graph-level encoding~\cite{KDZDS2017}.

 \subsubsection{Decoding}
 
Once all action embeddings $Z_i$ and state embedding $Z_s$ are computed in the~encoding stage, the~decoder will leverage these representations to compute approximated state-action value function $Q(s, a;\Theta)$ which predicts the~expected future rewards of taking action $a$ in state $s$, and following the~policy $\pi_{\Theta}$ till the~end. Specifically, we concatenate the~embeddings of state and action and use it as decoder input. In principle, any decoder architecture may be used. Here, we use a standard feed-forward neural network. Formally, the~decoding process can be written as:

\begin{equation}
    Q(s, a^{(i)}; \Theta) = \psi_{\Theta}(Z_s \oplus Z_i),
\end{equation}

\noindent where $\psi_{\Theta}$ is a dense feed-forward neural network.

\subsection{Training}

We train our model on randomly generated Chimera instances. We found that the~minimal viable size of the~training instance is $C_3$  (as a reminder, $C_3$ is Chimera architecture with nine unit cells arranged into a $3 \times 3$ grid, which gives us 72 spins). Smaller instances lack couplings between clusters, crucial in full Chimera, which leads to poor performance. We generate $\{ J_{i,j}\}$ and $\{ h_i\}$ from normal distribution $\mathcal{N}(0,1)$ and starting spin configuration $\sigma$ from uniform distribution. To introduce low-energy instances, we employed the~following pre-processing procedure. For each generated instance, with probability $p=10\%$, we perform standard simulated annealing before passing the~instance through SGNN. 

We seek to learn approximation of optimal action-value function $Q(a,s;\Theta)$, so as reinforcement learning algorithm we used standard n-step deep $Q$ learning~\cite{PW1994,DQL} with memory replay buffer. During episode we collect  sequence of states action and rewards $\tau = (s_0, a_0, r_0, \ldots, s_{T-1}, a_{T-1}, r_{T-1}, s_T)$ with terminal state as final element. From those we construct $n$-step transitions $\tau^{n}_{t} = (s_t, a_t, r_{t,t+n}, s_{t+n})$ which we collect in memory replay buffer $\mathcal{B}$. Here $r_{t,t+n} = \sum_{k =0}^{k=n} \gamma^{k} r_{t+k}$ is return after $n$-steps. 

\subsection{Simulated Annealing with Reinforcement}

Simulated annealing with reinforcement (SAwR) combines machine learning and classical optimization algorithm. Simulated annealing (SA) takes its name from a process in metallurgy involving heating a material and then slowly lowering the~temperature to decrease defects, thus minimizing the~system energy. In SA, we start in some state $s$ and in each step, we move to a randomly chosen neighboring state $s'$. If a move lowers energy $E(s)$ of the~system, we accept it. If it doesn't, we use the so-called the~Metropolis-Hasting criterion.

\begin{equation}
\mathbb{P}(\text{accept} \, s' ~\vert~ s) = \min (1, e^{-\beta \Delta E}),       
\end{equation}
 
\noindent where $\Delta E = E(s') - E(s)$ and $\beta$ denotes inverse temperature $1/T$. In our case, the~move is defined as a single-spin flip. Simulated annealing tends to accept all possible moves at high temperatures (i.e., lower $\beta$). However, it likely accepts only those moves that lower the~energy at low temperatures. 
 
 Our idea is to reinforce random sampling with a trained model. It means that at low temperatures, instead of using the~Metropolis-Hasting criterion, we perform a single pass of the~DIRAC episode.
 
\section{Experiments}

We collected data from the D-Wave 2000Q device using default parameters (number of samplings, annealing time, etc.). We have generated 500 random instances of sizes $C_4$, $C_8$, $C_{12}$ and $C_{16}$, corresponding to systems of sizes 128, 512, 1152 and 2048 spins respectively. We used identical distributions to training instances, so $\{ J_{i,j}\}$ and $\{ h_i\}$ was generated from normal distribution $\mathcal{N}(0,1)$ and starting spin configuration $\sigma$ from uniform distribution.
We then used quantum annealing to obtain the low energy states of generated instances. 

\begin{figure}
\centering
\begin{subfigure}{\textwidth}
    \includegraphics[width=\textwidth]{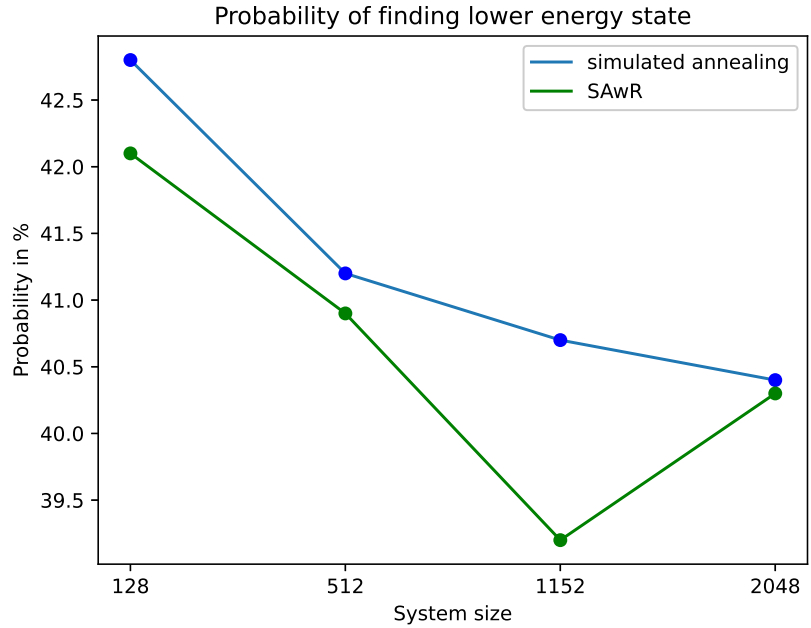}
    \caption{\small The probability of finding a lower energy state was computed over 500 random instances for each Chimera size. Here we decided to omit results for a single pass of the spin flipping procedure because its results were the order of magnitude worse, making the figure hard to read.}
    \label{a}
\end{subfigure}
\hfill
\begin{subfigure}{\textwidth}
    \includegraphics[width=\textwidth]{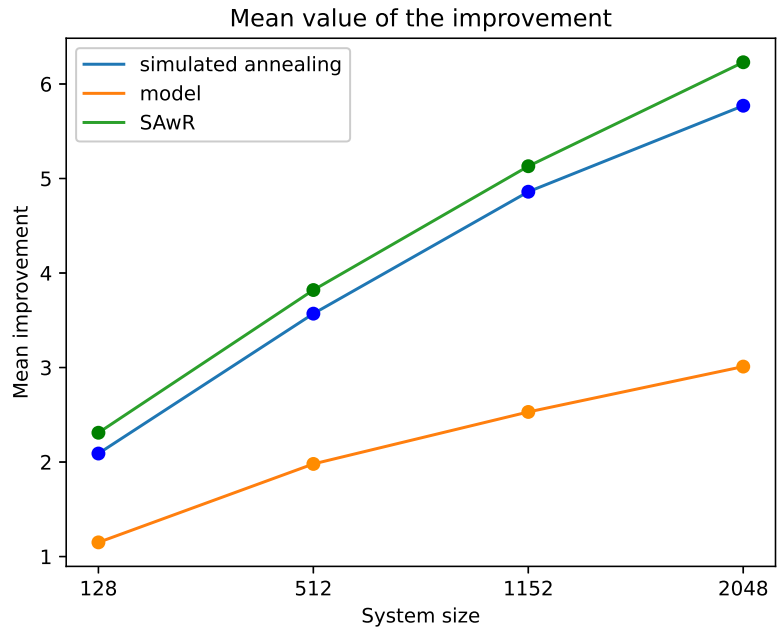}
    \caption{\small Mean value of the improvement. We define the value of the improvement as the difference between starting energy state and the lowest energy state found by the tested method in abstract units of energy. The mean was computed over all instances where the method could find improvement}
    \label{b}
\end{subfigure}
\caption{\small Results of our experiments for tested metrics. By model, we define a single pass of the spin flipping procedure. SAwR denotes simulated annealing with reinforcement.}
\label{fig:figures}
\end{figure}

We have used three methods - standard simulated annealing, a single pass of spin-flipping procedure and simulated annealing with reinforcement. Results are shown in figures \ref{a} and \ref{b}.

We tested for two metrics: the probability of finding lower energy states and the mean value of an improvement over starting energy state. To compute the probability for each Chimera size, we started with proposed solutions obtained from quantum annealer and tried to lower them using different tested methods. Then we counted those instances for which a lower energy state was found. We define the value of the improvement as the difference between starting energy state and the lowest energy state found by the tested method in abstract units of energy. 

Simulated annealing with reinforcement achieved lower probabilities of finding a lower energy state. Although the difference between SAwR and traditional simulated annealing is slight, its consistency across all sizes suggests that it is systemic rather than random noise. The single pass of the spin-flipping procedure was the order of magnitude worse, reaching approximately $1\%$ success rate.

It is interesting that, on average, SAwR was able to find a better low energy state than simulated annealing, but still, the difference is not significant.

\section{Discussion and Further Work}

The real-life Chimera graph is a much more complex problem than the~regular lattice employed in~\cite{FSNLSL2021}, which may be the reason for poor performance. However, we managed to replicate the excellent scaling of DIRAC. We trained our model on relatively small instances and employed it for large ones. We did not observe significant differences in performance relative to the size of the system. Much more work is needed because more complex architectures are deployed (Pegasus, Zephyr) by D-wave systems.  

 One possible avenue of research is changes in architecture. Right now goals of the~encoder and decoder are quite different. The~encoder tries to encode information while the~decoder leverages them for reinforcement learning. Training them both at the~same time might be difficult. One option is to divorce them from a single architecture. For example, we may use graph autoencoder~\cite{GAE,GAE2} to train the~encoder. Then for the~reinforcement learning part, we would use an already trained encoder and train only the~neural network responsible for approximating the~action-value function.
 
 Another option is to use different reinforcement learning algorithms. Asynchronous methods have been shown to consistently beat their synchronous counterparts, especially asynchronous advantage actor-critic~\cite{A3C}, but asynchronous Sarsa or Q-learning also look promising. Methods based on Monte Carlo tree search inspired by AlphaZero~\cite{AZ} also seems promising. It has shown excellent performance on tasks involving large search space (ex. Chess, Go).

\section*{Acknowledgments} 
This research was supported by the Foundation for Polish Science (FNP) under grant number TEAM NET POIR.04.04.00-00-17C1/18-00 (LP, ZP, and BG). TS acknowledges support from the National Science Centre (NCN), Poland, under SONATA BIS 10 project number 2020/38/E/ST3/00269. This research was partially funded by National Science Centre, Poland (grant no 2020/39/ B/ST6/01511 and 2018/31/N/ST6/02374) and Foundation for Polish Science (grant no POIR.04.04.00-00-14DE/ 18-00 carried out within the Team-Net program co-financed by the European Union under the European Regional Development Fund). For the purpose of Open Access, the author has applied a CC-BY public copyright license to any Author Accepted Manuscript (AAM) version arising from this submission.

\bibliographystyle{splncs04}
\bibliography{bibliography}

\end{document}